\documentclass[graybox]{svmult}

\usepackage{type1cm}        % activate if the above 3 fonts are
\usepackage{makeidx}         % allows index generation
\usepackage{graphicx}        % standard LaTeX graphics tool
\usepackage{multicol}        % used for the two-column index
\usepackage[bottom]{footmisc}% places footnotes at page bottom

\usepackage{newtxtext}       %
\usepackage{newtxmath}       % selects Times Roman as basic font

\makeindex             % used for the subject index
\begin{document}

\title*{Vector vortex solitons and soliton control in vertical-cavity surface-emitting lasers}
\author{T. Ackemann, T. Guillet, H. Pulham, G.-L. Oppo}
\institute{T. Ackemann \at SUPA and Department of Physics, University of Strathclyde, Glasgow G4 0NG, Scotland, UK, \email{thorsten.ackemann@strath.ac.uk}
\and T. Guillet \at Laboratoire Charles Coulomb (L2C), Univ. Montpellier, CNRS, Montpellier, France, \email{Thierry.Guillet@umontpellier.fr}
\and H. Pulham \at SUPA and Department of Physics, University of Strathclyde, Glasgow G4 0NG, Scotland, UK, \email{g.l.oppo@strath.ac.uk}
\and G.-L. Oppo \at SUPA and Department of Physics, University of Strathclyde, Glasgow G4 0NG, Scotland, UK, \email{harry.pulham@strath.ac.uk}
}
\maketitle

\abstract{The properties of vector vortex beams in vertical-cavity-surface emitting lasers with frequency-selective feedback is investigated. They are interpreted as high-order vortex solitons with a spatially non-uniform, but locally linear polarization state. In contrast to most schemes to obtain vector vortex beams relying on imprinting the polarization structure, vector vortex solitons form spontaneously due to the near polarization degeneracy in vertical-cavity devices. We observe radially, hyperbolic and spiral polarization configurations depending on small residual anisotropies in the system and multi-stability between different states. In addition, we demonstrate flip-flop operation of laser solitons via in principle local electronic nonlinearities. Combining the two themes might open up a route for a simple device enabling fast switching between different vector vortex beams for applications. The investigations connect nicely the fields of nonlinear science, singular optics, structured light and semiconductor laser technology.}

% 10-20 printed pages
% textwidth apparently 136 mm

%\begin{figure}[b]
%\sidecaption
% For example, with the graphicx style use
%\includegraphics[scale=.65]{figure}
%
% If no graphics program available, insert a blank space i.e. use
%\picplace{5cm}{2cm} % Give the correct figure height and width in cm
%
%\caption{If the width of the figure is less than 7.8 cm use the \texttt{sidecapion} command to flush the caption on the left side of the page. If the figure is positioned at the top of the page, align the sidecaption with %the top of the figure -- to achieve this you simply need to use the optional argument \texttt{[t]} with the \texttt{sidecaption} command}
%\end{figure}

\section{Introduction}
\label{sec:intro}
This paper will review and provide new results on \emph{high-order spatial lasers solitons} which have the non-trivial polarization characteristics of \emph{vector vortex beams}. They provide a fascinating bridge between nonlinear science and singular optics, both thriving research fields on their own.  Solitons or solitary waves are shape-stable waves for which the dispersive or diffractive spreading typical of linear waves is counteracted by nonlinearities. The first observation of a shape-stable `wave of translation' was done by the Scottish engineer Scott Russell in the Union Channel in 1834 \cite{russel1844}. In optical context, solitonic behaviour can occur in time (temporal soliton \cite{hasegawa73,mollenauer80}) or in space (spatial soliton \cite{chiao64,bjorkholm74}), both in the simplest case described by a Nonlinear Schr\"odinger Equation (NLSE)\cite{berge98}.  In the spatial case, self-localization can be understood by the concept of a self-induced nonlinear waveguide. The beam induces a refractive index distribution in a nonlinear medium with an intensity dependent refractive index, which -- for a medium in which the refractive index is increasing with intensity, i.e.\ a self-focusing medium -- will counteracts the linear diffractive spreading. For the soliton solution, the beam writing the refractive index distribution is the fundamental mode of the induced waveguide in a self-consistent way. If the concept of a single-humped fundamental soliton is already intriguing, it is the more remarkably that more complex wave distributions can propagate as high order solitons. Nevertheless, high-order fibre solitons were observed already in the first experiments of Mollenauer \cite{mollenauer80}. For spatial solitons, the analogy with linear modes suggests that a suitably nonlinear adaption of the ring-shaped doughnut as the first-order Laguerre-Gauss mode might be a suitable candidate for a high-order soliton. However, it turns out that their propagation is unstable resulting in the break-up to two fundamental solitons as the ring-shaped intensity structure is subject to a modulational instability \cite{rosanov94,tikhonenko95,firth97}
%\cite{rosanov94,tikhonenko95,firth97,skryabin98,desyatnikov05}
in the same way as a plane wave is modulationally unstable in a self-focusing medium \cite{berge98}.

One way to stabilize this high-order soliton is by adding dissipation and driving.  In optics, this is usually achieved by placing the medium into a cavity. The resulting states are an attractor of the dissipative dynamics and referred to as \emph{dissipative solitons, cavity solitons or localized structures} \cite{akhmediev05,akhmediev07,tlidi94,firth98}. Dissipative systems support a wider range of solitary waves or localized structures than the conservative NLSE. Examples are the stability of spatial solitons in two spatial dimensions in the Lugiato-Lefever equation \cite{lugiato87} (the dissipative extension of the NLSE describing a coherently driven cavity with an intra-cavity Kerr medium) \cite{firth96a}, the existence and stability of fundamental solitons in lasers \cite{fedorov91,rosanov92,bazhenov91} and coherently driven cavity systems \cite{firth96a} with absorptive media, and, relevant for this paper, the stabilization of vortex solitons in lasers. These were predicted in lasers with saturable absorbers \cite{fedorov03,rosanov05} and observed in coupled \emph{vertical-cavity surface-emitting lasers (VCSELs)}, one operated as a gain device, one as saturable absorber \cite{genevet10a}.
%\cite{genevet10a,genevet12}.
A VCSEL is a semiconductor laser in which the emission is in the direction of the epitaxial growth \cite{iga00} which allows for a large Fresnel number \cite{grabherr98,grabherr99} enabling self-organization and solitons independent from boundary conditions (see Sec.~\ref{sec:setup}). Vortices in self-focusing Ginzburg-Landau models were predicted in \cite{crasovan00,mihalache08a}. Refs.~\cite{paulau10,paulau11} predicted stable vortex solitons with a saturable or cubic (i.e.\ Kerr-like) self-focusing nonlinearity  coupled to an additional linear filter. This system provides a minimal model for a VCSEL with frequency-selective feedback \cite{tanguy08,tanguy08a,radwell09}. The experimental observation of vortex solitons in such a system was reported in \cite{jimenez13}.

These vortex solitons were homogeneously linearly polarized and can be described in a quasi-scalar theory. However, light is in general a vector wave providing many additional degrees of freedom for soliton formation. It is not possible to give a full account of the literature but we refer to \cite{haelterman94b,oppo99,gilles17}. Typically, these vector solitons consists of localized patches of one polarization state embedded in another one with a polarization domain wall in between. Interestingly, to our knowledge the first observation of stable high-order dissipative solitons were vector solitons in this sense \cite{pesch05}. What we are investigating here are structures in which the polarization is \emph{continuously spatially varying}.  They are usually referred to as \emph{`vector vortex beams (VBBs)'} \cite{milione11,souza14} and possess a circularly symmetric intensity structure combined with a spatially non-uniform polarization field and a polarization singularity. We will review their properties, generation and applications in Sec.~\ref{sec:vectorintro}. For the moment the important point to note is that there are usually created by a bespoke, potentially complex, setup imposing the polarization structure onto the beam (see Sec.~\ref{sec:vectorintro} for examples). In contrast, VVBs form spontaneously in the VCSEL with frequency-selective feedback discussed here \cite{jimenez17}. This is enabled by the high circular symmetry of a VCSEL in the transverse plane of the microcavity. This is known to allow a degeneracy or near-degeneracy of polarization states but investigations were focused on linearly polarized states and switching between them, see e.g.\  \cite{choquette94,sanmiguel95b,exter98b,sanmiguel00a,ackemann05b}. One paper predicted the possibility of VVBs in free-running VCSELs already in 1997 \cite{prati97} and thus established an early link between singular optics  and VCSEL technology, but, appearing before the great upsurge in interest in vector vortex states, did not obtain the attention due and was not backed up by experiments.

The main contribution of this paper (Sec.~\ref{sec:vector}) is to review the properties of VVBs and solitons in VCSELs with frequency-selective feedback, and to report on experimental progress and new insights compared to Ref.~\cite{jimenez17}. In particular, we will argue in Sec.~\ref{sec:interpret} why these states have been observed now in the VCSEL with feedback but not in free-running VCSELs as predicted in Ref.~\cite{prati97}. As a disclaimer, we would like to caution that we are using here the phrase vector vortex solitons for the vector vortex structures observed and we will argue in Sec.~\ref{sec:interpret} for this interpretation. However, to our knowledge, there is no detailed theory demonstrating their stability in VCSELs with frequency-selective feedback. We hope that this review might instigate theoretical investigations to this effect.  A recent prediction of vector vortex solitons in a laser with saturable absorption was made in Ref.~\cite{mayteevarunyoo18}. It is important to note that vector vortex solitons in single-pass propagation schemes are known to be unstable for self-focusing media \cite{ishaaya08}, similar to the quasi-scalar case \cite{rosanov94,tikhonenko95,firth97}, although less unstable than their quasi-scalar counterparts \cite{bouchard16}. Vector vortex solitons in single-pass propagation schemes are predicted to exist for self-defocusing nonlinear media \cite{ciattoni05}, but to our knowledge there is no experimental observation.

The final subject we are going to cover in this paper is the all-optical control of a laser soliton as a memory element \cite{firth96,brambilla96,spinelli98,barland02}. This property is intrinsically linked to dissipative soliton representing localized states. They can be present or absent under the same conditions and are hence necessarily bistable (see Sec.~\ref{sec:whyOB}). This makes optical solitons attractive for all-optical processing and storage schemes and motivated some of the research in semiconductor lasers \cite{spinelli98,barland02}. One of the advantages of laser cavity solitons is that they do not need to be sustained by a coherent holding beam with high spatial and temporal coherence but can draw their energy from an incoherent optical or electrical input. However, for the prospects of soliton control, this also removes an obvious and easy source for the control beams. We will review aspects of laser cavity soliton switching in Sec.~\ref{sec:history_switching} in more detail, but although already the first experiments on VCSEL lasers solitons demonstrated bistability and some control \cite{tanguy07,tanguy08,genevet08}, switch-on and switch-off of laser cavity solitons had been only obtained in situation involving thermal and/or non-local effects for one of the two directions of switching \cite{tanguy08,genevet08,elsass10}. In Sec.~\ref{sec:flipflop}, we will demonstrate flip-flop operation, i.e.\ the setting and re-setting of an optical memory element by external optical pulses directed directly on the memory element, using a two-colour control scheme exploiting only local electronic nonlinearities.

\section{Mechanism of bistability in lasers with frequency-selective feedback}\label{sec:whyOB}
\begin{figure}[htb]
\includegraphics[width=75mm]{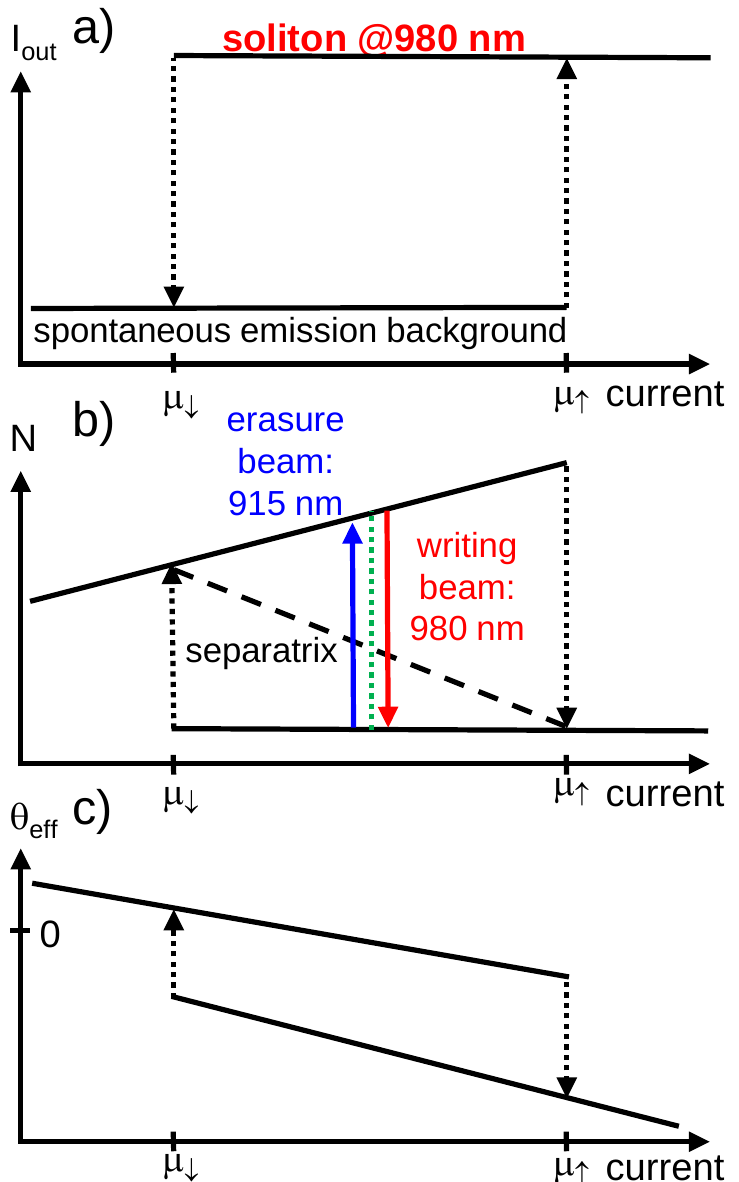}
\sidecaption\caption{\label{fig:why_ob} Schematic drawings of hysteresis loops (solid black lines) for optical intensity (a), carrier density (b) and effective detuning $\Theta_{\rm eff}$ (c) if the injection current is slowly increased and decreased. Panel b) illustrates also the unstable branch (dashed black line) working as separatrix. Somewhere around the centre of the hysteresis loop there is the Maxwell point (dashed green line) at which the on- and off-states of the soliton have equal stability. Flip-flop operation in Sec.~\ref{sec:flipflop} is obtained around this point. Injected pulses around 980~nm (red line) will be amplified, decrease carrier density and hence work as writing beam. Injected pulses around 915~nm (blue line) increase carrier density via optical pumping and hence work as erasure beam.  As the detuning varies (c), the carrier density is not totally constant even under lasing condition (as assumed in b), but this is a quantitative, not qualitative consideration. }
\end{figure}

The experiment relies on the interaction of VCSEL with a frequency-selective element. A VCSEL is a semiconductor laser based on a high-Finesse plano-planar microcavity, e.g. \cite{iga00}. The technical details are presented in Sec.~\ref{sec:setup}. The cavity is very short (on the order of 1.2~$\mu$m) and hence it runs in single longitudinal mode, but the transverse aperture can be very large (about 200~$\mu$m) \cite{grabherr98,grabherr99} so that it has a very high Fresnel number and can run in many transverse modes or, in the present context, form self-localized structures which are independent from each other.  The basic observation is illustrated in Fig.~\ref{fig:why_ob}a and an experimental example is given in the inset of Fig.~\ref{fig:setup_flipflop}. If the current is slowly increased from zero, first only low-amplitude spontaneous emission is observed until there is a an abrupt transition to a localized high-amplitude state, the soliton. The soliton stays for a certain range of current, if the current is increased further, but more importantly it also stays on if the current is decreased again below the switch-on point $\mu_\uparrow$. Switch-down takes place only at a lower current $\mu_\downarrow$ creating a hysteresis loop. Within this hysteresis loop the soliton can be manipulated by external control beams that will be the subject of Sec.~\ref{sec:flipflop}.

Fig.~\ref{fig:why_ob}b indicates what is happening with the carrier density at the position of the soliton during the process. Increasing the current in the non-lasing situation increases the carrier density. At the switch-on threshold there is a sudden drop of carrier density due to the onset of stimulated emission. Beyond threshold, for the operating laser soliton, carrier density is clamped and will stay approximately constant over the existence range. At the switch-off point, the carrier density is abruptly switching back to the unsaturated value. A free-running laser, in contrast to coherently driven passive cavity or optical feedback systems \cite{kreuzer90,spinelli98,kreuzer96,schaepers99}, amplifiers \cite{barland02} or lasers with injection \cite{hachair04}, does not require an external holding beam. On the one hand, this is a major advantage, on the other hand this implies that there is also no option to derive control beams from the holding beam and to implement coherent soliton control via the optical hysteresis loop. In contrast, soliton control takes place via the carrier density as will be discussed in detail in Sec.~\ref{sec:flipflop}.

For an understanding of the origin of the hysteresis loop it is important to realize that the main effect of the current sweeping the hysteresis loop is not changing the gain but the detuning condition between the VCSEL cavity resonance $\omega_c$ and the VBG resonance $\omega_{\rm VBG}$, at least once sufficient gain is achieved to allow lasing for overlapped or close resonances. For consistency with the terminology used in \cite{scroggie09,jimenez16}, we define
\begin{equation}\label{eq:Theta_def}
\Theta := \frac{\omega_c - \omega_{\rm VBG}}{\kappa}= \frac{\omega_c}{\kappa} =\Theta_0-\zeta\mu,
\end{equation}
where $\kappa$ is the cavity linewidth (HWHM) for the field, $ \omega_{\rm VBG}=0$ is used as reference frequency, $\Theta_0$ is the detuning for zero current and $\zeta$ is the proportionality factor between current and resonance shift due to Ohmic heating ($\zeta=5.23$ in \cite{jimenez16}). This is the `cold' cavity resonance. In a semiconductor laser, the refractive index depends on carrier density $N$ and in first approximation this phase-amplitude coupling is described by a simple proportionality constant, the so-called Henry's alpha-factor $\alpha$ \cite{henry82}. Hence we define an effective detuning
\begin{equation}\label{eq:Theta_eff}
\Theta_{\rm eff} := \Theta_0-\zeta\mu+\alpha N,
\end{equation}
which is depicted in Fig.~\ref{fig:why_ob}c. For zero current, the cavity resonance is blue-detuned to the VBG resonance, $\Theta_{\rm eff} >0$. Increasing the current red-shifts the cavity resonance via the heating and blue-shifts it via the carrier injection, but the former effect is prevailing (\cite{jimenez16} assumed $\alpha=5$ as a reasonable value for the alpha-factor) and the detuning is decreasing. If the detuning is sufficiently small, the reduced losses will favour an increase of (amplified) spontaneous emission, which in turn will decrease carrier density, increase refractive index and hence further decrease the detuning leading to a further decrease of losses and hence a tendency to higher intensity. At the spontaneous switch-on point, the non-lasing state becomes unstable to fluctuations due to this positive feedback and the system switches abruptly to an high amplitude state in which the resonances of VCSEL and VBG are (approximately) aligned. This high-amplitude state can be maintained to some extent (till the switch-off point) even if the cold cavity detuning is increased again via a reduction of current. This means that the laser with frequency-selective feedback works via a form of dispersive optical bistability \cite{lugiato84}.  The stability of the non-lasing state surrounding the soliton is achieved because it is off-resonant \cite{paulau11,firth10} and not because it has higher losses as in lasers with saturable absorbers \cite{fedorov91,rosanov92,bazhenov91}. Laser solitons in semiconductor-based devices were realized first  in \cite{tanguy08} based on the dispersive effect. The counterpart relying on absorptive optical bistability can be realized by coupling two VCSELs face-to-face \cite{genevet08}, one more strongly pumped than the other, although dispersive components might be important in that configuration also.

The cold cavity resonance offset $\Theta_0$ can be adjusted by adjusting the ambient temperature of the device. Hence a situation can be realized in which effective zero detuning is obtained for elevated temperature and low currents. In this situation, the first structure appearing at threshold is the fundamental soliton \cite{radwell09,jimenez16}. This is the situation investigated in Sec.~\ref{sec:flipflop}. Alternatively, at low ambient temperature a much higher Ohmic heating and hence current is needed to reach the effective zero detuning condition. In this situation, the spontaneous switch-on is to more complicated structures and in particular high-order solitons as vortices and vector vortices \cite{jimenez16,jimenez17}. As high threshold current implies high threshold gain, this tendency towards high-order structures is reasonable. This is the situation to be investigated in Sec.~\ref{sec:vector}.

However, before turning to this subject, it should be mentioned as a final caveat that in an ideal, homogeneous system there would not be a spontaneous switch-on to a soliton but to a spatially extended modulated state, a spatial pattern, as the system would remain in the non-lasing or homogeneous state until the threshold for modulational instability (MI). Bistable solitons exist below the MI point down to the switch-off point. For experimental realizations using semiconductor microcavities this ideal situation cannot be realized in spite of these microcavities being marvels of vacuum-deposition technology as a growth error of a single monolayer is enough to change the detuning significantly due to the high cavity Finesse. Hence the solitons are pinned to certain positions in which the detuning conditions are favorable \cite{barland02}. For the situation considered here these are the most red-shifted part of the cavity as investigated in detail in \cite{ackemann12}. However, a huge body of experimental and numerical investigations indicate that within the hysteresis loop the solitons keep their important solitonic features except the mobility to move freely within the transverse aperture of the device and hence the community regards them as solitons \cite{barland02,ackemann09a,barbay11}.

\begin{figure}[hbt]
\includegraphics[width=70mm]{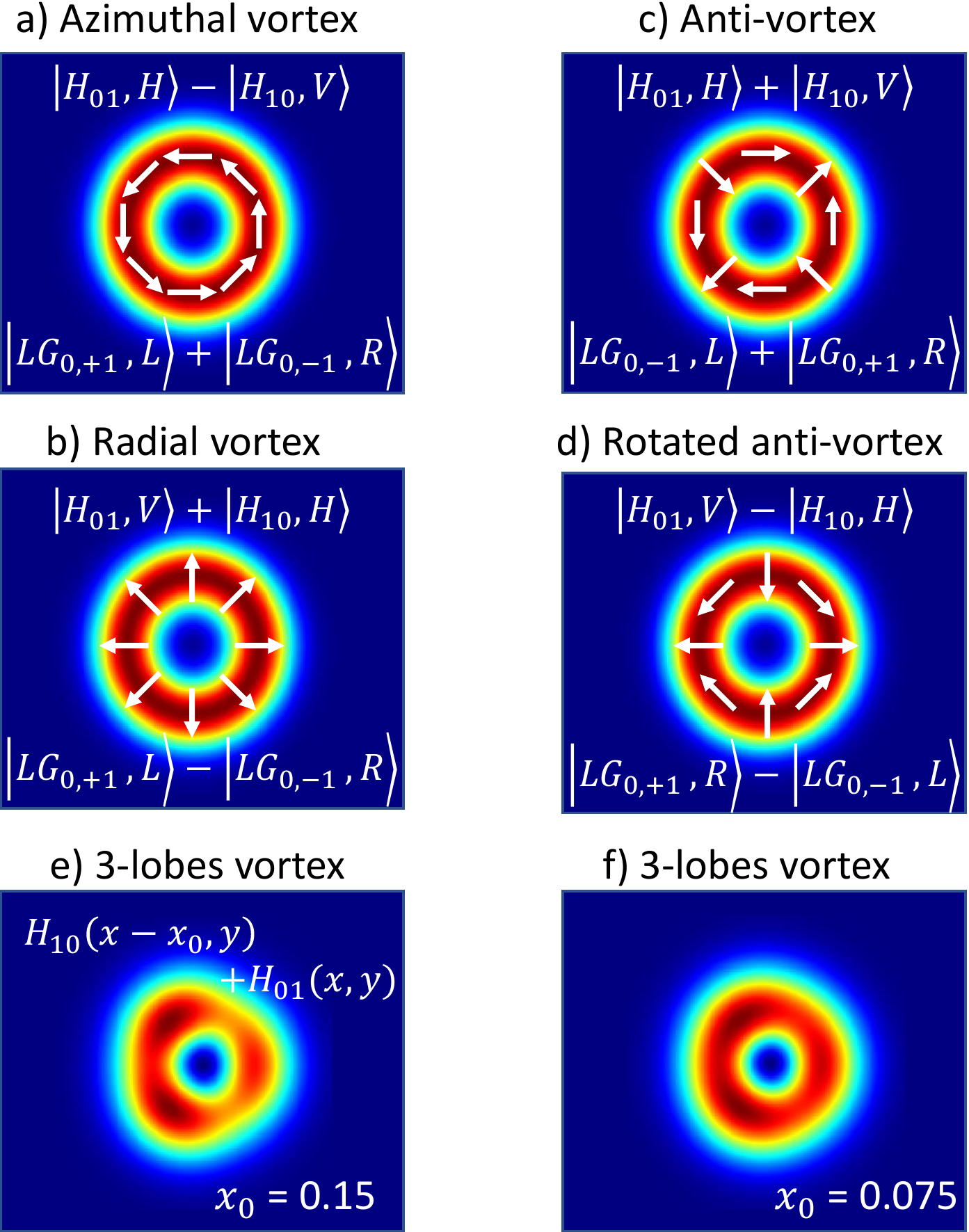}
\sidecaption\caption{\label{fig:vectorvortex}a-d) Illustration of different vector vortex beams as obtained from different superpositions of orthogonally polarized first order in a linearly polarized basis (or of circularly polarized Laguerre -Gaussian modes). Intensity in pseudocolour scale ranging dark blue (low intensity) to dark red (high intensity). e, f) If the centre of the two orthogonal Hermite modes are slightly displaced by a fraction of $x_0$ of the beam waist of the fundamental mode, the ring is perturbed to a three spot structure.}
\end{figure}

\section{Vector vortex solitons} \label{sec:vector}
\subsection{What are vector vortex beams?} \label{sec:vectorintro}
Vector vortex beams are a subset of beams with complex intensity, phase and polarization structure investigated in the field of `structured light' \cite{rubinsztein-dunlop17,rosales18,forbes19,otte20}. In contrast to `full
Poincar\'e beams' utilizing the full Poincar\'e sphere \cite{beckley10}, \emph{`vector vortex beams'} (VVBs) \cite{milione11,souza14,forbes16} have a spatially non-uniform, but locally linear, polarization field. Although in general other configurations are possible, the beams investigated in practice possess a circular symmetric intensity structure in form of a doughnut. Fig.~\ref{fig:vectorvortex} shows the states relevant for our considerations. For these first order VVBs four basis modes are needed. The first option is to construct them from the two first-order Hermite-Gaussian modes ($H_{10}$, $H_{01}$) with horizontal (H) and vertical (V) polarization. Figs.~\ref{fig:vectorvortex}a, b show the azimuthally and radially polarized states, where the polarization structure possesses cylindrical symmetry  (`cylindrical vector beams' \cite{zhan09,milione11}). Beams with hyperbolic polarization structure are referred to as `anti-vortices' \cite{maurer07} or `$\pi$-vortices' \cite{milione11}. They exist in two versions (Figs.~\ref{fig:vectorvortex}c, d) with principal axes rotated by 45$^\circ$. These beams do not have a phase singularity in the centre (an unique optical phase is ill defined with two polarization components anyway) and hence do not carry orbital angular momentum, but they contain a singularity in the polarization field, respectively in the relative phase between the polarization components. Any projection onto a linearly polarized state yields a first-order Hermite-Gaussian modes with an orientation depending on the direction of the projection (see Sec.~\ref{sec:observations}).

An alternative construction can be done in the Gaussian-Laguerre basis $LG_{0,\pm 1}$ and using the circular polarization basis (L, R). In this representation, it is apparent that VVBs contain orbital and spin angular momentum in a correlated manner. After projection on circular polarization basis, the resulting fields carry orbital angular momentum.

VBBs have intrinsic appeal and beauty, provide novel fundamental aspects in quantum optics and enable new applications or enhance existing  ones in engineering and science. Examples are  tight focusing \cite{dorn03}, micro-machining \cite{allegre12}, optical trapping \cite{kozawa10,roxworthy10,otte20}, simultaneous spectroscopy of multiple polarization channels \cite{fatemi11} and beam transformation in nanophotonics \cite{tischler14}. Particularly exciting is the realization that the correlation between spatial and polarization degrees of freedom resembles entanglement and might open up novel schemes for the use in quantum optics \cite{qian11,ndagano17,qian17} and sensing \cite{berg-Johansen15}.

Hence, considerable effort was spent on creating these unusual polarization states \cite{zhan09,rosales18,wang20}, relying on a substantial engineering effort based on specialized equipment as tailored laser resonators \cite{fridman08,senatsky12,ngcobo13,naidoo16}, meta-surfaces and spatially varying wave plates \cite{machavariani07,liu14,bauer15}, Mach-Zehnder interferometers \cite{roxworthy10}, modal control in few-mode fibers \cite{ramachandran13,ndagano15}, spatial light modulators \cite{maurer07,forbes19,otte20}, tailored Fresnel reflection from glass cones \cite{radwell16},  and  polariton microcavities \cite{lagoudakis09,manni13,sala15}. In contrast, we demonstrate the spontaneous emergence of these structures in a conceptually simple system, a highly symmetrical vertical-cavity surface-emitting lasers (VCSEL) with frequency-selective feedback \cite{jimenez17}. To our knowledge, this was the first experimental observation of the spontaneous formation of a vector vortex beam from spontaneous symmetry breaking. A more recent observation in a polariton laser is described in Ref.~\cite{hu20}.

As a final remark before turning to the experiment, one can perturb the intensity distribution of VVB without destroying the polarization structure by providing a slight offset between the centres of the two orthogonal Hermite-Gaussian modes forming the VVB. In that case the intensity along doughnut ring becomes modulated (Fig.~\ref{fig:vectorvortex}f) and becomes a `3-spot structure' at larger shifts  (Fig.~\ref{fig:vectorvortex}e).

\subsection{Experimental setup}\label{sec:setup}
The setup for the cavity soliton laser and the VCSEL devices used are described in detail in \cite{grabherr98,grabherr99,radwell09,schulz-ruhtenberg09} and reviewed in \cite{ackemann13,ackemann16}. The VCSELs used are broad-area electrically pumped devices. Three InGaAs quantum wells are serving
as gain medium leading to emission in the 980~nm range. The quantum wells are surrounded by passive
AlGaAs spacer layers with a total thickness of one wavelength. The cavity is closed by high reflectivity
distributed Bragg reflectors (DBR) with 33 layers AlGaAs/GaAs on the top side (p-contact) and 22
layers on the bottom side (n-contact). The laser has an emission wavelength around $975~{\rm nm}$ at room temperature. The emission takes place through the n-doped Bragg reflector
and through the transparent substrate. In this so-called bottom-emitting geometry a reasonable uniformity
of carrier injection can be achieved over fairly large apertures. A 200~$\mu$m diameter circular oxide aperture provides optical and current guiding. This active diameter
is much larger than the effective cavity length of about 1.2~$\mu$m. As a result, the VCSEL has a
large Fresnel number allowing for the formation of many transverse cavity modes of fairly high order and of solitons which are independent of boundary conditions.

\begin{figure}[hbt]\begin{center}
\includegraphics[width=100mm]{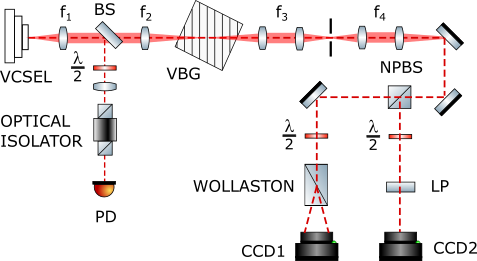}
\end{center}
\caption{\label{fig:setup_Stokes} Setup for measuring spatially resolved Stokes parameter in a VCSEL with frequency-selective feedback. $f_i$: lenses, VBG: volume Bragg grating (the inscribed grating is tilted by an angle $<1^\circ$ with respect to the surface in reality), BS: beam sampler, NPBS: non-polarizing beam splitter, $\frac{\lambda}{2}$: half-wave plates, PD: amplified, slow photodiode (in reality there is a whole detection branch here but only the photodiode is used in these investigations), LP: linear polarizer, CCD: charge-coupled device cameras. The two folding mirror on the right are included for compactness of the drawing. The actual beam path is straight.}
\end{figure}

The optical setup is illustrated in Fig.~\ref{fig:setup_Stokes}. Frequency-selective feedback is provided by an external volume Bragg grating (VBG). The VBG has a reflection peak of about 95\% at 978~nm and a bandwidth of slightly larger than 0.1~nm (FWHM). The VCSEL is collimated by $f_1=8~{\rm mm}$ focal length plano-convex aspheric lens. The second lens is a $f_2=50~{\rm mm}$ focal length plano-convex lens and is used to focus the light onto the VBG. The lenses are arranged as an afocal telescope giving a $6.25:1$ magnification factor onto the VBG. This external cavity is self-imaging, i.e.\ every point of the VCSEL is imaged at
the same spatial position after each round trip therefore maintaining the high Fresnel number of the VCSEL cavity and ensuring local feedback compatible with self-localization.

In the standard experimental scheme the output is monitored via the Fresnel reflection from a wedged intra-cavity beam sampler BS (front uncoated, back AR-coated).  The cavity is isolated from the detection setup by an optical isolator. In this experiment, it is mainly used for monitoring the light-current (LI) characteristic by an amplified, slow photodetector. In addition, the output intensity distributions are monitored for alignment and analysis by charge-coupled device cameras  (CCD) after suitable optics, in near field and far field (CCD1, CCD2 depicted in Fig.~\ref{fig:setup_flipflop}). The outcoupling is relying on Fresnel reflection and therefore is polarization dependent. The reflectivity is on the order of 10\% for s-polarized light and 1\% for p-polarized light and hence the polarization state of the intra-cavity is not adequately represented in this detector arm.

Hence, for the polarization resolved measurements presented here, we are analyzing the spatially resolved Stokes parameter in the field transmitted by the VBG. As the VBG is used at nearly normal incidence, the polarization state of the transmitted field is preserved. The image of the soliton on the VBG facet is relayed by a telescope ($f=150$~mm for both lenses) to an intermediate image plane. In this plane, a part of the VCSEL aperture can be selected for analysis via a movable circular aperture. This plane is imaged by a further telescope (typically $f=50$~mm for both lenses, but the second one can be adapted to serve different imaging needs) onto CCD cameras. The beam path is split by a non-polarizing beam splitter (Coherent, $T=0.33$, $R=0.67$). The transmitted arm is folded by a highly reflective mirror and traverses a half-wave plate (HWP) and Wollaston prism  before creating an image on a CCD-camera. This allows for the simultaneous monitoring of the linear polarizations components at $0^\circ$ and $90^\circ$ to calculate the spatially resolved Stokes parameter $S_1$ or the linear polarizations components at $+45^\circ$ and $-45^\circ$ to calculate the Stokes parameter $S_2$. In the reflected arm, there is either an amplified, slow photodetector  to monitor the polarization resolved total power or, more often, another CCD-camera monitoring the polarization at $+45^\circ$. In that case, both Stokes parameters $S_1$ and $S_2$ can be measured simultaneously, which is not only convenient but sometimes important, if structures change quite rapidly with changing current. As there is some jitter between different realizations of the experiment (see, e.g.\ Fig.~\ref{fig:OB_radial_lin}), combining the measurements from different runs to calculate Stokes parameters introduces potentially some artifacts. If both arms in the VBG branch are used with CCD-cameras, the LI-curve of the system is monitored via the intra-cavity beam sampler (a rough correction can be done for the anisotropic Fresnel reflection). For the experiments reported in Fig.~\ref{fig:interferogram}, a Mach-Zehnder interferometer is introduced into the beam path to confirm the phase properties of the vector vortex beams.

It should be noted that the intra-cavity beams sampler preserves the polarization state in transmission (0.99 vs.\ 0.9, ratio 1.1)  much better than in reflection (0.1 vs.\ 0.01, ratio 10). Nevertheless, it will introduce a dichroism into the system, which breaks the (nominal) polarization degeneracy and is hence expected to influence the formation of vector vortex beams. Hence, in some experiments, the intra-cavity beam sampler is removed and detection takes place only via the VBG.

% Harry, 170619 17.5deg
\begin{figure}[!hb]
\begin{center}
\includegraphics[width=110mm]{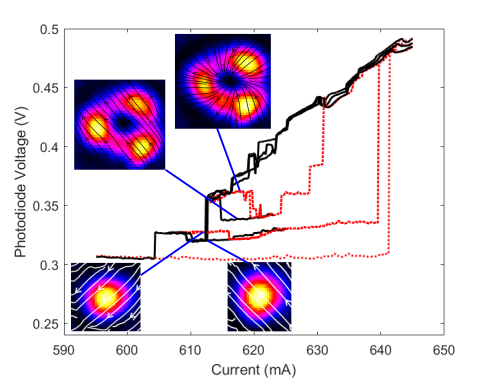}
\end{center}
\caption{\label{fig:LI} LI-curve projected on a suitable, but arbitrary linear polarization state chosen to have good discrimination between structures. Up-scans of current are denoted by red dotted lines, down-scans by solid, black lines. We did not distinguish the down scans by different colours because we do not believe that the variations contain useful information but that the ensemble demonstrates robust behaviour.  Insets show total intensity $S_0$ of structures in pseudo-colour (from low intensity black via blue and red to high intensity yellow-white) with polarization streamlines reconstructed from the polarization resolved Stokes parameters superimposed. Temperature of VCSEL submount: 17.5$^\circ$C.}
\end{figure}

\subsection{Principle observations}\label{sec:observations}
Figure~\ref{fig:LI} gives an overview on the scenario for the formation of vector vortex solitons. We report on new and more complete results obtained without a beam sampler in the external cavity. However, the principal behaviour is identical to the situation with this beam sampler reported in Fig.~2 of \cite{jimenez17}. Increasing the current, the output remains low, on spontaneous emission level, till there is a sudden switch-on to a high-amplitude lasing structure. This occurs at a quite large current of 641~mA since for the VCSEL submount temperature most of the thermal tuning to match VCSEL and VBG resonance comes from Ohmic heating (see discussion of Eq.~(\ref{eq:Theta_def})). Hence, the gain at threshold is quite large and fairly complex extended structures form which are not shown here but are reported in Fig.8 of \cite{jimenez16} and Fig.~2 of \cite{jimenez17}. Reducing the current again, the output power reduces, mainly as the lasing amplitude reduces, but partially also because the size of the lasing area reduces. However, typically a change of size and complexity of the lasing structures involves an abrupt sudden transition in modal shape and emitted total power. At a level of about 616~mA one reaches the vortex states. We will discuss these later.

Reducing the current further, the system switches at 613~mA to a single humped peak, the fundamental soliton, which is linearly polarized. The fundamental soliton shows a spontaneous switch to the orthogonal polarization at 609~mA (see also \cite{rodriguez17}). Polarization switching is a typical behaviour observed in VCSELs \cite{choquette94,sanmiguel95b,exter98b,sanmiguel00a,ackemann05b}. Further reducing the current, the fundamental soliton switches off at 604~mA. If the down-scan is stopped at 605~mA, and the current is increased again, one moves along the fundamental branch (with a polarization switch at 616~mA) until a switch to a very-high amplitude state occurs at 640~mA. If the scan is stopped before and reversed, the hysteresis loop of the polarization switch of the fundamental soliton can be explored.

If the original down-scan is stopped at 616~mA, the spatial structure of the total intensity, i.e.\ $S_0$ encountered has an approximate ring-shape with three dominant peaks along the ring. The polarization streamlines obtained from measuring the polarization resolved Stokes parameters are superimposed on the intensity image and show a radial polarization structure. This constitutes a radially polarized vector vortex beam where the deformation of the ring to three spots can be explained by spatial disorder in the VCSEL resonance (see discussion of Fig.~\ref{fig:vectorvortex}). Increasing the current, there is an abrupt transition (at 619~mA) to a vortex structure with nearly the same intensity distribution but it is now linearly polarized. These structures were identified as vortex solitons in \cite{jimenez13}. Reversing the current sweep, coexistence between the linearly and the radially polarized vortex is demonstrated. Increasing the current further, the system switches from vortex states to larger states at 629~mA and to even larger structures at 631~mA.

The final switch-up occurs to roughly the same branch for all situations, i.e.\ from the non-lasing state, the fundamental soliton branch, and, although typically via intermediaries, from the vortex branch. We kept all three realizations from the down-scan from these complex states in the figure to illustrate on the one hand robustness of the phenomena but also the importance of fluctuations. The main issues are thermal fluctuations and drifts leading to a variation of switch-on and switch-off points and a slight variations of amplitudes. Small spikes and jumps within one branch correspond usually to longitudinal mode hopes. We gave concrete numbers for the switching points in the discussion above in order to allow an easy identification of the points in the figure, but one should not put to much emphasis on the concrete numbers. However, we stress that the overall scenario is robust. In particular, there is an amazing amount of multi-stability in the system. For example, from 613-618 mA the off-state, the fundamental solitons, the vortex solitons and an even more complex state coexist, possibly even in several polarization configurations.

\begin{figure}[b]
\begin{center}
\includegraphics[width=100mm]{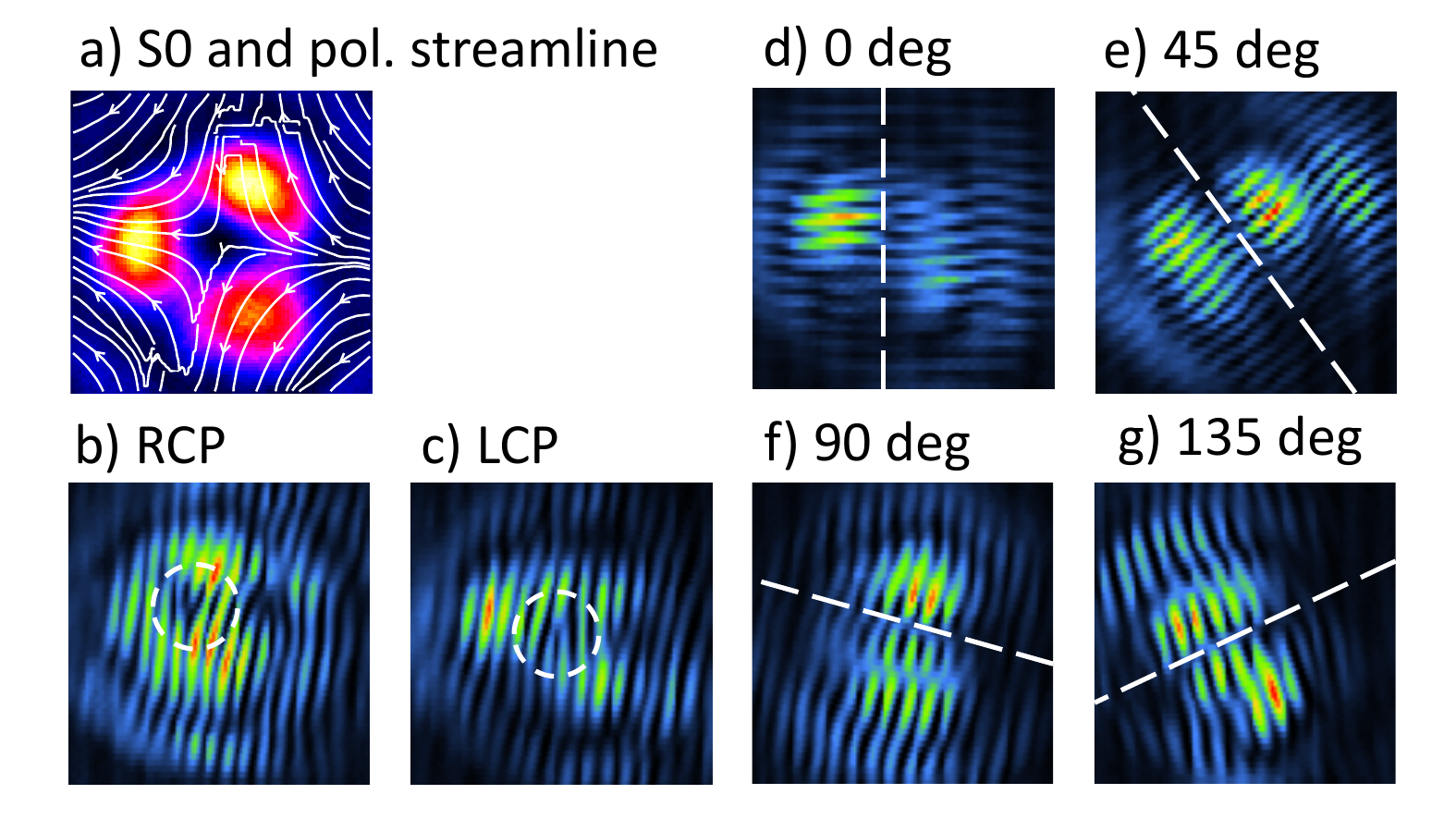}
\end{center}
\caption{\label{fig:interferogram}Mach-Zehnder interferometry of an anti-vortex: a) Total intensity $S_0$ (pseudo colours), and polarization streamline. b,c) Polarization-resolved interferograms projected on circular polarization states (assignment of RCP and LCP arbitrary). Dashed circles indicate the forks evidencing the phase singularity. It should be noted that we are looking at the self-interference of two doughnut beams and not of one doughnut beam with a plane reference wave. Hence two forks indicate the presence of only one phase singularity. d)-g) Polarization-resolved interferograms projected on linear polarization states in 45$^\circ$ steps. Dashed lines indicate the nodal line evidencing the $\pi$ phase shift.
}
\end{figure}

For the situation with the intra-cavity beam sampler placed within the cavity, the dominantly observed vector vortex is the anti-vortex with a hyperbolic polarization structure (see Fig.~\ref{fig:interferogram}a). LI-curves and further details are reported in Ref.~\cite{jimenez17}. Its phase structure is analyzed further in Fig.~\ref{fig:interferogram}. In subfigures d-g) the VVB is projected on linear polarization states and then interfered with itself in a Mach-Zehnder interferometer. After projection, the intensity structure is dominated by two elongated lobes. In between these lobes there is a nodal line with a $\pi$-phase jump across it as demonstrated by the staggered discontinuities of the interference fringes. This implies that the observed vector vortex beams are very close to the `ideal' ones whose construction from first order Hermite-Gaussian modes is reported in Fig.~\ref{fig:vectorvortex}. Obviously, there are differences from pure Hermite-Gaussian modes in the wings of the beams. Inspection of Figs.~\ref{fig:interferogram}d and f yields that the vortex under study is a superposition of a horizontally polarized $H_{10}$ and a vertically polarized $H_{01}$-mode. In order to obtain the hyperbolic polarization structure aligned (roughly) to the horizontal and vertical axes (see Fig.~\ref{fig:vectorvortex}d)) the relative phase between them needs to be $\pi$. Projection on polarization states at $\pm 45^\circ$ (Figs.~\ref{fig:interferogram}e and g) yields first order Hermite modes with the nodal axis orthogonal to the polarization axis as expected for the anti-vortex. The supplementary material \cite{jimenez17ss} of \cite{jimenez17} contains a movie which shows this continuous counter-rotation of the spatial structure, if the angle of the polarizer analyzer is rotated, clearly.

After projection on a circular polarization state (Figs.~\ref{fig:interferogram}b and c), the intensity structures are approximately circular and the phase structure contains now forks, i.e.\ evidence of phase singularities and not nodal lines. These are of opposite sign (opposite branching direction in the interferograms) for the two circular polarizations.  This reveals that the constituent modes are doughnuts themselves, i.e.\ carry orbital angular momentum, as expected if one calculates the corresponding transformations (e.g.\ \cite{manni13}). VVBs carry spin and orbital momentum degrees of freedom in a correlated manner.

\subsection{Complex hysteresis loops}
\begin{figure}[hbt]
% Harry, 170616, series 11, 19.5 deg, Harry 170616, a larger field of view is in ForThorsten.pptx, p. 5
\includegraphics[width=77mm]{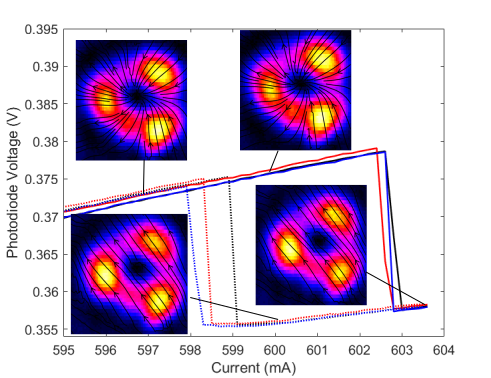}
\sidecaption\caption{\label{fig:OB_radial_lin} LI-curves and inset illustrating bistability between the radially and uniformingly linearly polarized vortex. The four realizations shown give an indication of the stability of the system on the time scale of several minutes. Temperature of VCSEL submount: 19.5$^\circ$C.}
\end{figure}

After having established the principal observations in the previous section, we return to a more detail of the system without the intra-cavity beam sampler in Fig.~\ref{fig:OB_radial_lin}. It shows the hysteresis loop between the radially polarized vortex and the linear polarized one, if the current is only swept up and down in a small vicinity of the transition. Repeated measurements done over a time scale of a few minutes show a robust coexistence on the one hand and some jitter of switching points and slight changes of amplitude. These are partially due to technical noise as current fluctuations and mirror vibrations, but mainly due to thermal fluctuations influencing the detuning between VCSEL and VBG being the most important and sensitive control parameter (see the discussion on Fig.~\ref{fig:why_ob}c). In particular, the switching point will be influenced by where the comb of the external cavity modes is with respect to the VBG and VCSEL resonances as this will render the spectrum of favourable soliton states discontinuous.  Thermal drifts are limiting the long term stability.

\begin{figure}[hbt]% 27/9/2016, series 10
\begin{center}
\includegraphics[width=0.95\textwidth]{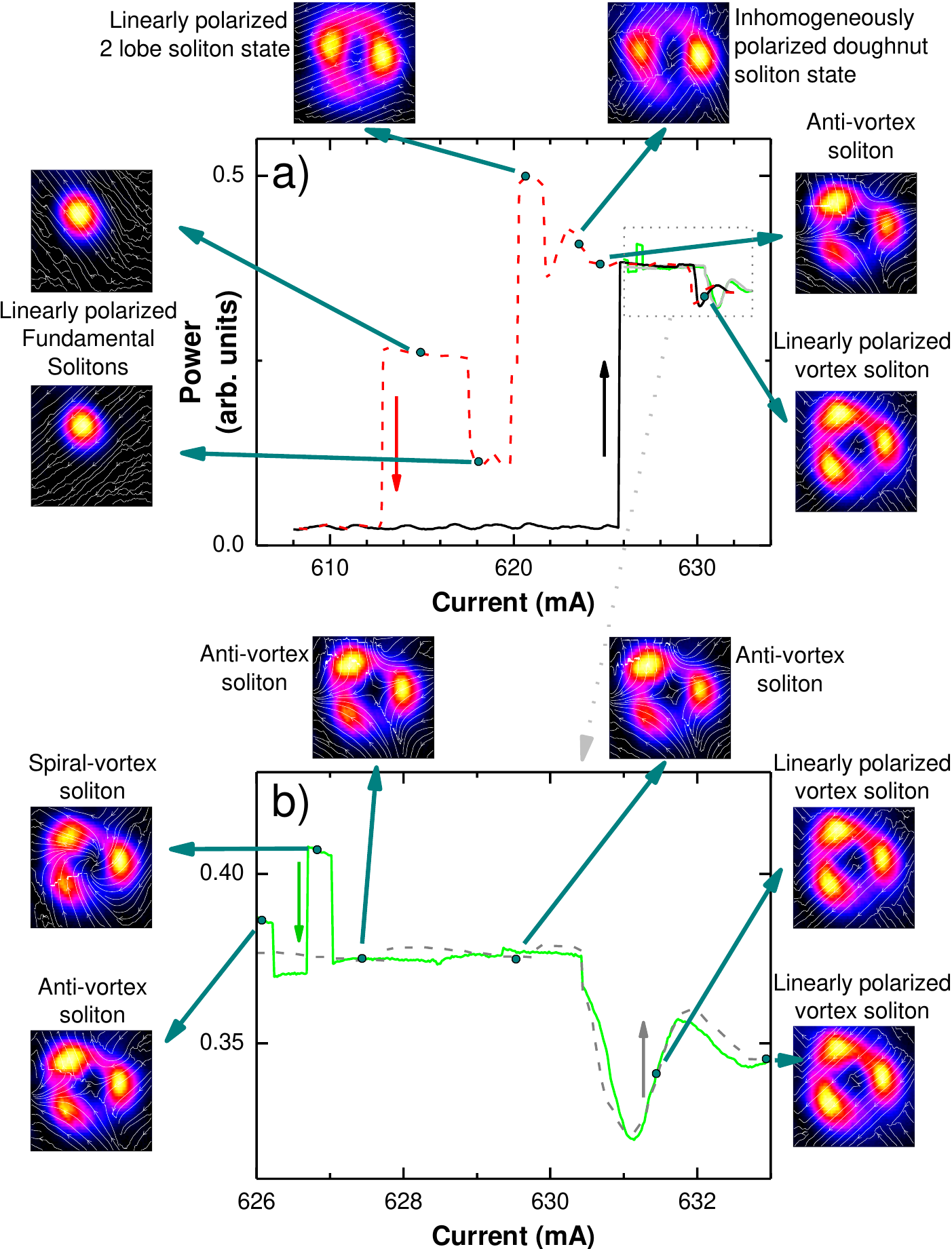}
\end{center}
\caption{\label{fig:withBS_largeloop} a) LI-characteristic of system with intra-cavity beam sampler with typical structures observed. Black line: up-scan, red dashed line: down scan at 0.1 mA per step and a waiting time of 500 ms waiting time per step. The dashed grey line and green solid are another realization of an up-scan and down-scan covering only the high current area at 0.02~mA per step and 500 ms waiting time per step. b) Blow-up of region beyond 626 mA.  Temperature of VCSEL submount: 19.6$^\circ$C.}
\end{figure}

Fig.~\ref{fig:withBS_largeloop} provides an example of the complex switching behaviour possible, now again for the system containing the intra-cavity beam sampler. Fig.~\ref{fig:withBS_largeloop}a is a large scale view. Switch-on from the off-state occurs here directly to the vector vortex states. This behaviour is observed frequently, although the more typical behaviour is the switch-on to larger states at even higher currents (Fig.~\ref{fig:LI} and Fig.~2 of \cite{jimenez17}). This behaviour might be related to minute misalignments of the retro-reflection angle of the VBG. Reducing the current one obtains a two-lobe structure. The fact that the laser does not seem to find a doughnut (or three-spot) shaped locked state but remains spatially symmetry broken might support the assumption of a small misalignment. This structure then switches to the fundamental soliton which in turn undergoes polarization switching. Interestingly, the two-lobe structure exists also in two polarization configurations. It can be linearly polarized (at about 621~mA) or consists of four domains of which the two adjacent have approximately orthogonal polarization and the opposing ones the same polarization (at 624~mA). The polarization principle axes of these domains are very similar to the ones of the two fundamental soliton ones with orthogonal polarization. This is not a vector vortex state as the polarization direction is not varying gradually over the beam. Instead, it seems to be related to the well-known polarization bistability of VCSELs along their principal axis but here realized in a spatially varying manner. It should be noted that also the approximately doughnut shaped three-spot structure can exist with such a polarization structure (Fig.~3 of \cite{jimenez17},  \cite{jimenez17ss} and inset of Fig.~\ref{fig:waveplate}a below). It is also interesting to note that the transition between the vector vortex beams and this structure can be gradual in the resolution of the experiment, whereas all transitions between vector vortex beams and from vector vortex to homogeneously linearly polarized ones are abrupt.

Fig.~\ref{fig:withBS_largeloop}b concentrates on the smaller range of currents above 626 mA. Between 626 mA and 630.5 mA the dominant structure is the anti-vortex. A 630.5 mA there is an abrupt transition to the linearly polarized vortex. In this case the hysteresis width is smaller than the jitter. For higher currents the dominant structure is the linearly polarized vortex switching between different longitudinal modes and slightly different polarization directions. In the lower part, on the backward scan, a further novel structure is encountered, a vortex with a spiral polarization structure. It can be thought of as a superposition of radially and azimuthally polarized vortices or of all four Hermite-Gaussian basis modes (Fig.~\ref{fig:vectorvortex}a,b). It exists only in a very small range and not necessarily in all scans (e.g.\ it is in the green, but not the red-dashed down-scan in Fig.~\ref{fig:withBS_largeloop}b). The difference between the two realizations here is the scan speed. The spiral vortex occurs only for small enough current steps and can be easily missed.
    %It is also interesting to note that it did not occur at all in the first experimental campaign leading to the discovery of the anti-vortex as the first spontaneous vector vortex structure.
It is typically embedded in the existence range of the anti-vortex (see green curve in \ref{fig:withBS_largeloop}b). Minute adjustments, i.e.\ much smaller ones that the angular width of the vector vortex beams, influence whether the spiral vortex -- and on a more relaxed level the other vortices -- can be obtained.  This indicates that the vector vortices are degenerate in a perfectly symmetric system but that small anisotropies and/or fluctuations select between them.

\subsection{Influencing polarization selection by intra-cavity waveplates}
\begin{figure}\begin{center}
\includegraphics[width=85mm]{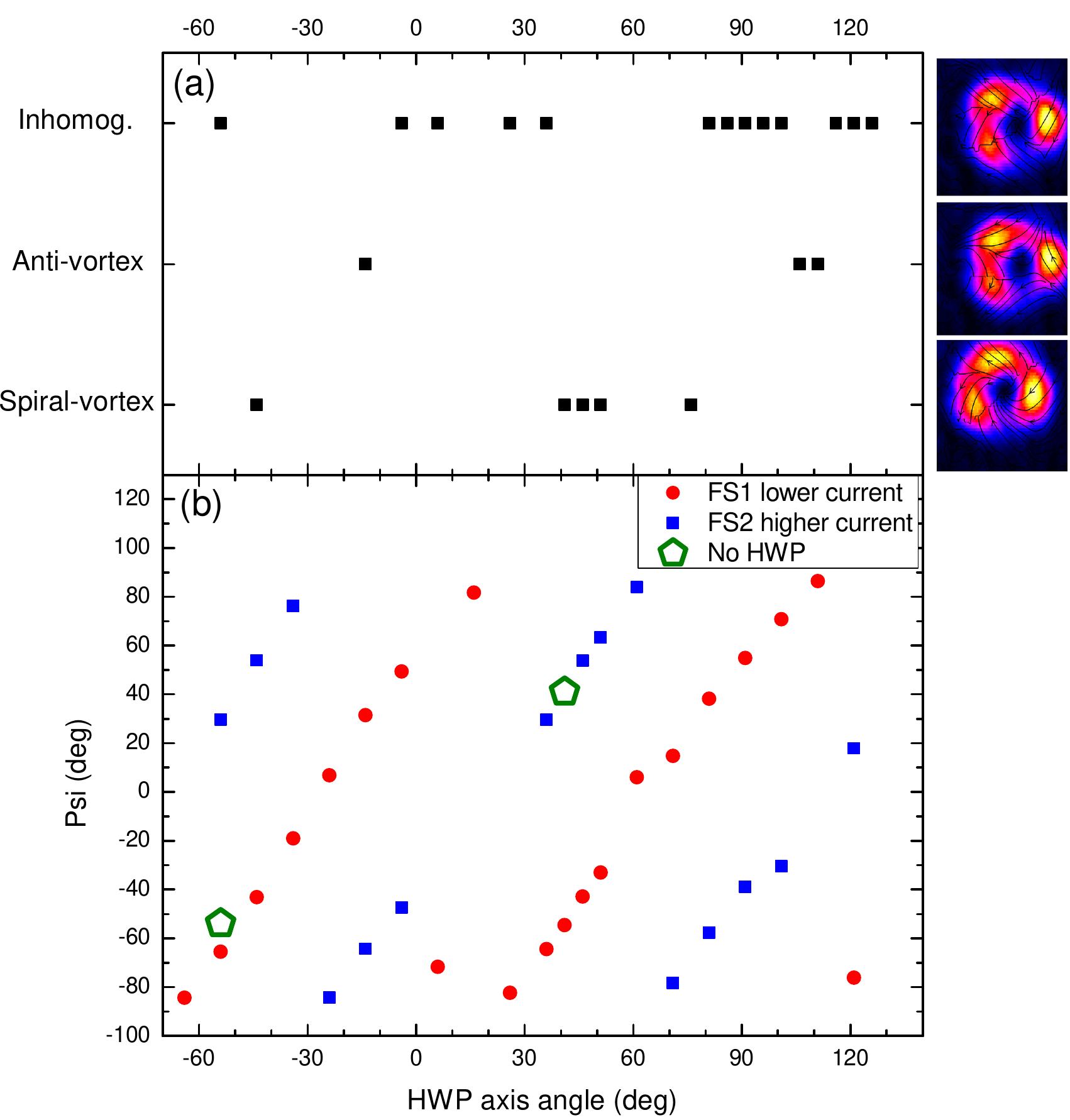}
\end{center}
\caption{\label{fig:waveplate} Influence of an intra-cavity HWP on polarization direction of fundamental solitons $\psi$ (b) and selection of VVB or domain structures (a). The reference axis for the angles is vertical. a) Indication of which type of VVB is observed, insets show $S_0$ and polarization streamlines. We report the VVB obtained in the current range before switch-down to the fundamental soliton occurs, but checks were carried out via additional, smaller LI-curve, whether other VBBs might be accessible from other initial conditions.
b) Red dots denote the main polarization direction of the fundamental soliton occurred at lower currents, blue of the one at higher current. The green rhombus indicates the selected polarization without HWP. The slope of polarization angle vs.\ HWP angle is 2 as a HWP turns the polarization by two times the deviation from its principal axis. The BS is present in the cavity.
}
\end{figure}

Although desirable, it is not straightforward to include means for controlling weak anisotropies in our setup, but we can introduce a strong anisotropy via a half-wave plate(HWP). This is introduced into the external cavity between the telescope lenses. First, we investigated the influence on the polarization properties of the fundamental solitons (Fig.~\ref{fig:waveplate}b). It is obvious that the direction of polarization of the fundamental solitons is rotated in the same way as the HWP is rotated, which can be expected for such a large phase anisotropy like the HWP. Over most of the range both polarization states can be obtained with the exception of a region centred around a polarization angle $\psi\approx -10^\circ$, which is interestingly roughly in between the two polarization state of the two fundamental solitons without the introduction of the HWP. This seems to be the least favourable orientation with only the intrinsic anisotropies, and hence only the favoured polarization state might survive.

%Feb 2017
Fig.~\ref{fig:waveplate}a summarizes the observations at what angle of the HWP which type of VVB or domain structure is observed. In addition, linearly polarized vortex beams have been observed along some point of the hysteresis loop at most angles of the HWP. It is hard to recognize a pattern behind the selection but it is interesting to note that only a single type of VBB was found for each angle.  The second observation is that the spiral vortex is a robust feature at the appropriate anisotropy, whereas it had been very difficult to obtain in the experiments discussed before. At the angles where no VBB or domain structure is reported, the linearly polarized vortex has a vertical polarization i.e.\ orthogonal to the horizontal one for which only one fundamental soliton exists. This reenforces the impression that vertical/horizontal polarization corresponds to some maximum polarization anisotropy. Apart from these observations, the figure mainly supports the notion that all or at least many kind of VVB exist in the system and the anisotropies decide whether are stable and accessible via LI-curves.  We will discuss ideas how to develop these investigations in Sec.~\ref{sec:conclusion}.

\subsection{Interpretation}\label{sec:interpret}
From the observation reported, it appears that small anisotropies in the system decide which VVB is stable and can be observed. This is in line with the theoretical analysis in \cite{prati97} for a free running VCSEL predicting the existence and stability of the three types of VVB that we observe (anti-vortex, spiral vortex and radially polarized vortex). Perfect cylindrical symmetry is best for a large existence range, but they survive modest anisotropies. No detailed analysis of their stability against linear polarized vortices or between different kind of vector vortex beams is however given in Ref.~\cite{prati97}. It might be even possible that a VVB we did not observe is stable but cannot be accessed by conventional LI-curves.

The VVB states share many properties with high-order dissipative solitons as the  localization to a small region in a broadly pumped plano-planar cavity, their coexistence with the non-lasing zero background and the abrupt emergence. The induced waveguide should be similar to the linearly polarized ones investigated in \cite{jimenez13} as it depends on total intensity $S_0$. For these, theoretical investigations support the soliton interpretation \cite{paulau10,paulau11}. This supports the notion that the solitonic behaviour comes from quasi-scalar guiding ($S_0$) and the polarization structure forms spontaneously under the conditions of polarization degeneracy or near-degeneracy. The deformation of the doughnut rings to 3-spots is no counter-argument against solitonic character, although poorly understood. Corresponding generalized vortex solitons were predicted \cite{desyatnikov05a} and indications observed \cite{minovich09} in single-pass conservative systems and termed azimuthons. Theoretical predictions exists also for dissipative systems \cite{fedorov03,soto-crespo09} including a simplified model for a VCSEL with frequency-selective feedback \cite{paulau15p}. A detailed theoretical treatment is currently beyond our resources but we hope that this contributions triggers theoretical efforts to this effect.

Coming back to why VVB as predicted by \cite{prati97} were not observed in free-running VCSELs, it is important to realize that in all real VCSEL the polarization degeneracy will be lifted to some extend by the electro-optical effect in electrical pumped devices \cite{exter97} and elasto-optical effects introduced via uncontrolled strain in the growth process \cite{doorn96}. This leads to birefringence and possibly dichroism. In addition, even monolayer fluctuations will influence detuning conditions. Typically these fluctuations will not have rotational symmetry, thus lifting the degeneracy between $H_{10}-$ and $H_{01}-$modes (or better to say between nonlinear counterparts for the soliton states). Thus it might be that locking to a VVB is not possible. However, even minute tilts of the VBG will influence the feedback phase significantly and in a slightly different way across an extended structure like a doughnut. We demonstrated frequency tuning of fundamental solitons and differential frequency detuning between spatially separated solitons before \cite{paulau12,ackemann13}. Hence, we conclude that minute alignment changes of the VBG, which are much smaller than the angular width of the soliton, can provide the necessary frequency fine tuning to allow the compensation of small intrinsic anisotropies and the formation of VVB. Tilting the VBG away from the condition used will destroy these states and favour linearly polarized ones or irregular domain structures. On the other hand, we stress that the observation of VVB is robust. Otherwise, we would not have discovered them as the observation reported in \cite{jimenez17} was not anticipated.
%\begin{figure}[b]
%\includegraphics[width=100mm]{interpretation}
%\caption{\label{fig:interpretation}...}
%\end{figure}

\section{Flip-flop operation of laser cavity solitons} \label{sec:flipflop}
\subsection{Soliton control in systems with and without holding beams} \label{sec:history_switching}
As indicated in the introduction, a major advantage of a cavity soliton laser is that it does not need a broad coherent driving (or holding beam, HB) but can draw all its energy out of incoherent pumping. However, for soliton manipulation, the holding beam provides actually the nice feature that a beam can be split off easily forming a focused addressing beam (AB) which can be used to switch solitons up and down in a conceptually simple way \cite{firth96,spinelli98,barland02}. By controlling the phase between HB and AB and thus locally the driving intensity, the system can be switched from one of the bistable states to the other, up for constructive and down for destructive interference. In contrast, there is no beam before a cavity soliton switches on, but switch-on can be initiated by an external beam in the vicinity of the soliton frequency (ignition beam, IB). This has been achieved in various VCSEL laser systems relying on dispersive \cite{tanguy08,tanguy07,radwell10} or absorptive nonlinearities \cite{bache05,genevet08,elsass10} or possibly a combination of them \cite{genevet08}. Inspection of Fig.~\ref{fig:why_ob}b yields that the switch-on is related to a drop in carrier density which then shifts the detuning via phase-amplitude coupling. Similarly, in the system based on saturable absorption \cite{genevet08}, reduction of carrier density caused by the IB leads to an increase in cavity finesse and hence soliton switch-on. For both systems, an erasure would demand adding carriers which cannot be achieved with optical injection close to the laser frequency but demands optical pumping at higher energies at which the semiconductor is absorptive. Such an experiment has not been performed for solitons in electrically pumped VCSELs, yet, but the switch-off of solitons has been obtained in \cite{tanguy08,radwell09,genevet08} with an erasure beam (EB) aimed at the side of the solitons. This method utilizes that the detuning conditions vary over the cross-section of the wafer and hence of the broad-area laser. The EB drags the solitons out of their preferred positions in the centre of the traps providing the best detuning condition. They do not recover, if the EB is switched off. However, a two-colour flip-flop operation scheme has been demonstrated before for bistable states in edge-emitting semiconductor amplifiers \cite{maywar00}. Interesting experiments have been also performed using a mode-locked TiSa-laser in the absorptive regime of VCSEL amplifiers \cite{barbay06,barbay07} and VCSEL with integrated saturable absorber \cite{elsass10} demonstrating flip-flop operation with a single colour providing optical pumping. Here the switch-down can be understood via the mechanism discussed above but the up-switch is not fitting the expectation. The up-switch involves substantial delays and it is argued in \cite{barbay06,barbay07,elsass10} that the possibility of up-switching is due to a combination of thermal and non-local effects. Flip-flop operation relies on hitting a sweet spot in parameter space. Hence we are addressing below the question of robust two-colour flip-flop operation of laser cavity solitons.

\subsection{Experimental setup} \label{sec:flipflopsetup}
\begin{figure}[b]\begin{center}
\includegraphics[width=100mm]{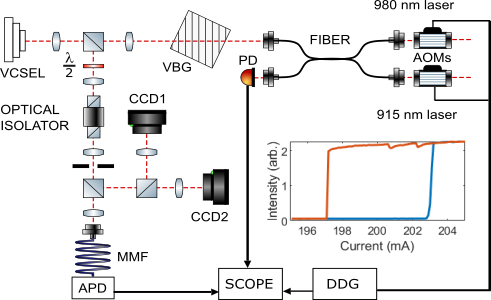}
\end{center}
\caption{\label{fig:setup_flipflop} Setup for flip-flop operation of laser cavity solitons. AOM: Acousto-optical modulator, MMF: multi-mode fibre, DDG: digital delay generator, APD: avalanche photodiode, CCD1(2): charge-coupled device cameras monitoring near and far field of the VCSEL.VBG: volume Bragg grating. The inset shows a LI-curve obtained at a VCSEL temperature of 47$^\circ$C for the fundamental soliton.}
\end{figure}

The optical setup is illustrated in Fig.~\ref{fig:setup_flipflop}. A laser at 980 nm is used for the IB beam and a laser at 915 nm for the EB. Acousto-optical modulators (AOM) allow for a fast pulsing of the beams (minimal pulse duration about 25~ns for the IB, 200~ns (100~ns with a strong reduction of efficiency) for the EB). The beams are launched into single-mode fibres and combined in a $2\times2$ fibre-coupler with a nominal coupling rate of 50\%. The output of one arms is monitored by a DC-coupled photodetector with 50~MHz bandwidth. The beam from the other output is collimated and then focused onto the VBG, size-matched to the soliton image at the VBG and hence to the soliton in the VCSEL. The beam is then aligned to be on axis and hitting the soliton via the cameras monitoring near and far field. The output of the VCSEL is monitored by an AC-coupled avalanche photodiode (APD) operated in the linear regime with a bandwidth of 1.7~GHz. For some measurements, it was replaced by a DC-coupled photodetector with 50~MHz bandwidth. A delayed switching sequence of the AOMs  was coordinated via a digital delay generator. Peak powers after the fibre are 2 mW for the 980 nm laser and 14~mW for the 915 nm laser.

The submount temperature of the VCSEL was put to 40-47$^\circ$C. As a result, the detuning position for soliton switching (Fig.~\ref{fig:why_ob}c) is met at considerably lower currents than discussed in the previous sections and the first structure appearing is a fundamental soliton \cite{jimenez16}. Its hysteresis loop is depicted in the inset of Fig.~\ref{fig:setup_flipflop}.

\subsection{Experimental results}
\begin{figure}[b]
\includegraphics[width=77mm]{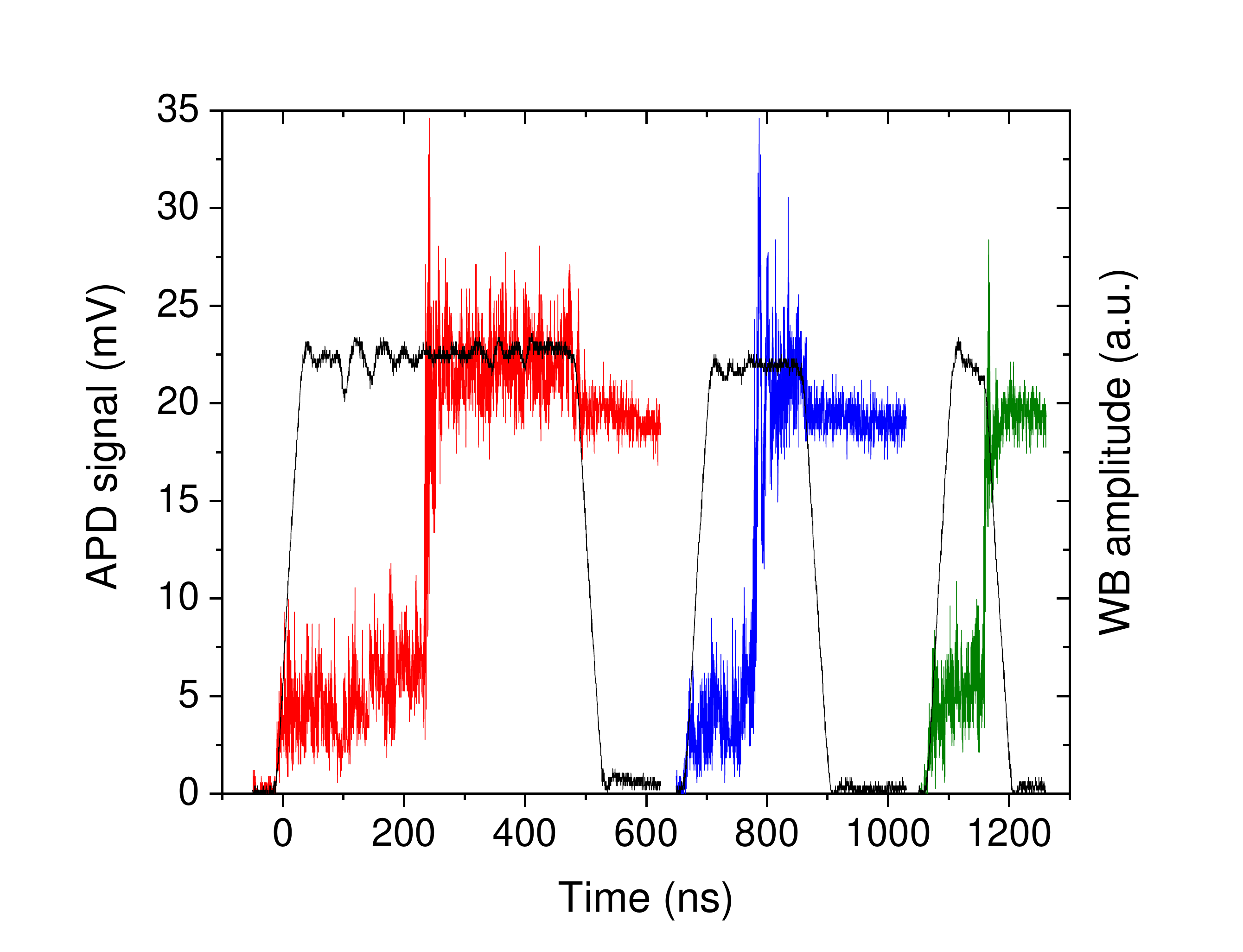}
\sidecaption\caption{\label{fig:switch-on} Three independently performed switch-on events (offset on time axis for clarity), with pulse widths of 500, 200, and 100 ns from left to right. The right-axis gives the WB amplitude (black lines) of the three pulses detected in the 50 MHz bandwidth photodiode. The left axis is the signal from the APD (coloured lines).  Temperature of VCSEL submount:  40$^\circ$C.}
\end{figure}

In a first experiment, the laser is biased within the hysteresis loop, but close (`detuning' 1 mA) to the spontaneous switch-on point. Fig.~\ref{fig:switch-on} demonstrates successful switch-on for pulse lengths of 100, 200 and 500 ns. The initial peak signal is from the IB itself. After a delay between 80 ns and 200 ns, there is a very fast and abrupt switch-on of the soliton. This stays on after the IB is switched off.
 The delay increases with increasing detuning from the spontaneous switch-on point and longer pulses are required (at constant amplitude, e.g.\ $\approx 500$~ns at $\approx 2$~mA, tens of microseconds at $\approx 3$~mA detuning). The jitter of the delay is related to fluctuations. These results match the observations and their numerical reproduction discussed in \cite{radwell10} and can be interpreted by the necessity to pass the separatrix indicated in Fig.~\ref{fig:why_ob}b in Sec.~\ref{sec:whyOB}. The separatrix  represents the unstable soliton solution separating the basins of attraction of the two stable states.

\begin{figure}[b]
\includegraphics[width=77mm]{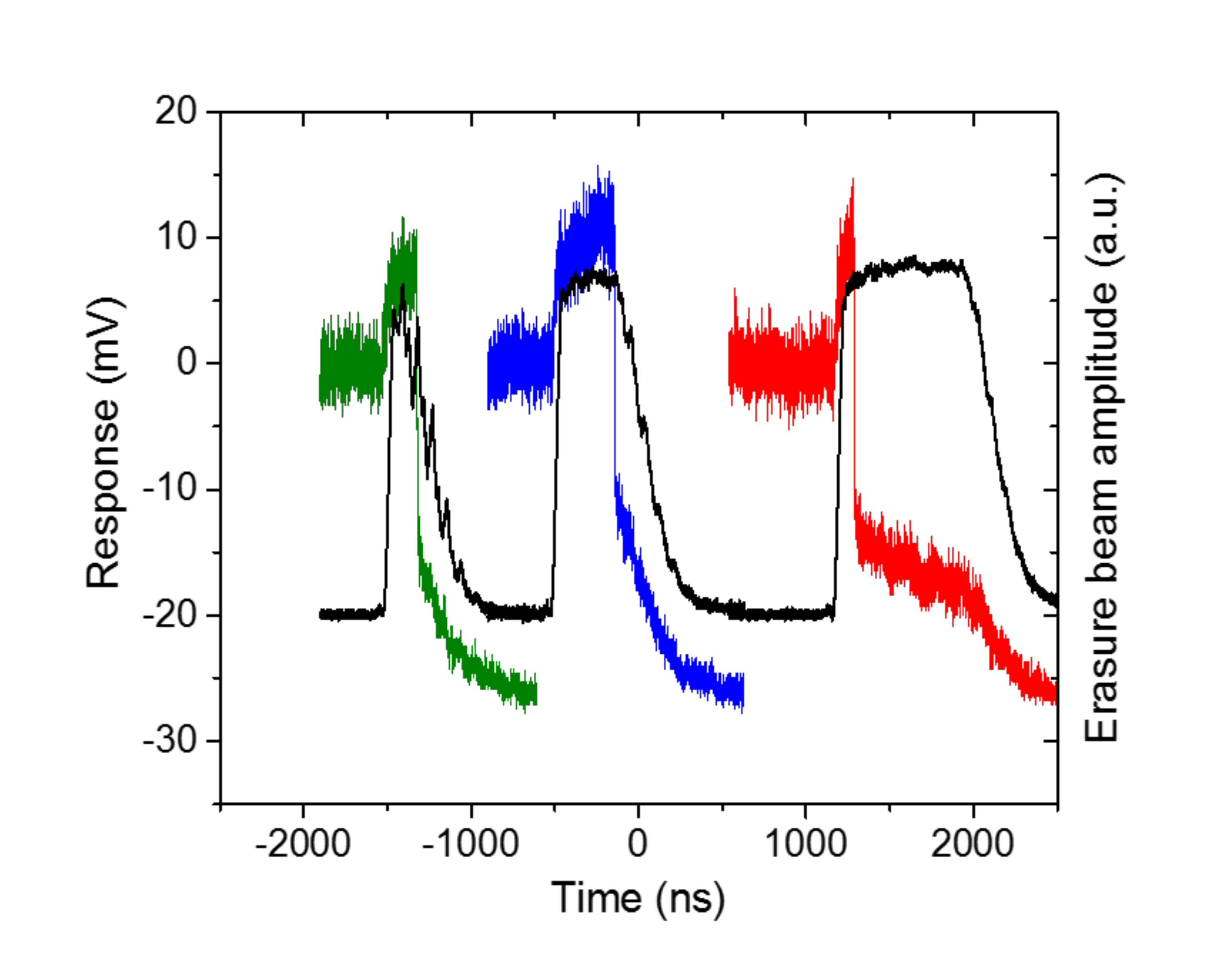}
\sidecaption\caption{\label{fig:switch-off} Three independently observed switch-off events, with  pulse widths of 200, 500, 1000 ns from left to right (see Fig.~\ref{fig:switch-on} for explanations).
    %The right axis shows input pulse amplitude (black) detected in the 50 MHz bandwidth photodetector. The left axis gives the signal from the (AC-coupled) APD.
    Note that, as the APD is AC-coupled, initially the soliton is on even if the signal is zero, which can be confirmed from examining the bias monitor output.}
\end{figure}

Switch-off events close to the spontaneous switch-off point are documented in Fig.~\ref{fig:switch-off}. For this, the  current is increased until spontaneous switch-on occurs, and then reduced close to the point of spontaneous switch-off. The initial increase in the detector signal is due to the detection of the EB. The fast reduction of signal is then due to the switch-off of the soliton, which occurs within the duration of the EB. The final tail is the decay of the EB. As for the switch-on, the minimum pulse width needed to induce switching at constant amplitude increases with increasing detuning to the microsecond range (at $\approx 3$~m A).

The scheme in Fig.~\ref{fig:why_ob}b indicates that somewhere typically close to the centre of the hysteresis curve, upper and lower state have an equal stability and roughly equal distance to the separatrix. This so-called    Maxwell point is hence the point to aim for flip-flop operation.  As time scales are then in the tens of microseconds range the AC-coupled APD is not working well for monitoring the signal. The DC monitor provides evidence of flip-flop operation but due to its low bandwidth demands millisecond pulses. Hence a DC coupled detector with 50 MHz bandwidth but much lower amplification is used. Fig.~\ref{fig:why_ob}b shows a realisation of a flip-flop event with input pulses of 50~$\mu$s width. The soliton switches on with the arrival of the 980 nm IB (it is unclear why the IB did not register in the detector in these experiments) and stays on until it is hit by the 915~nm EB. Further experiments indicate that the minimum pulse duration needed at the switching amplitudes available can be 10~$\mu$s, but switching is no longer robust.

\begin{figure}[b]
\includegraphics[width=77mm]{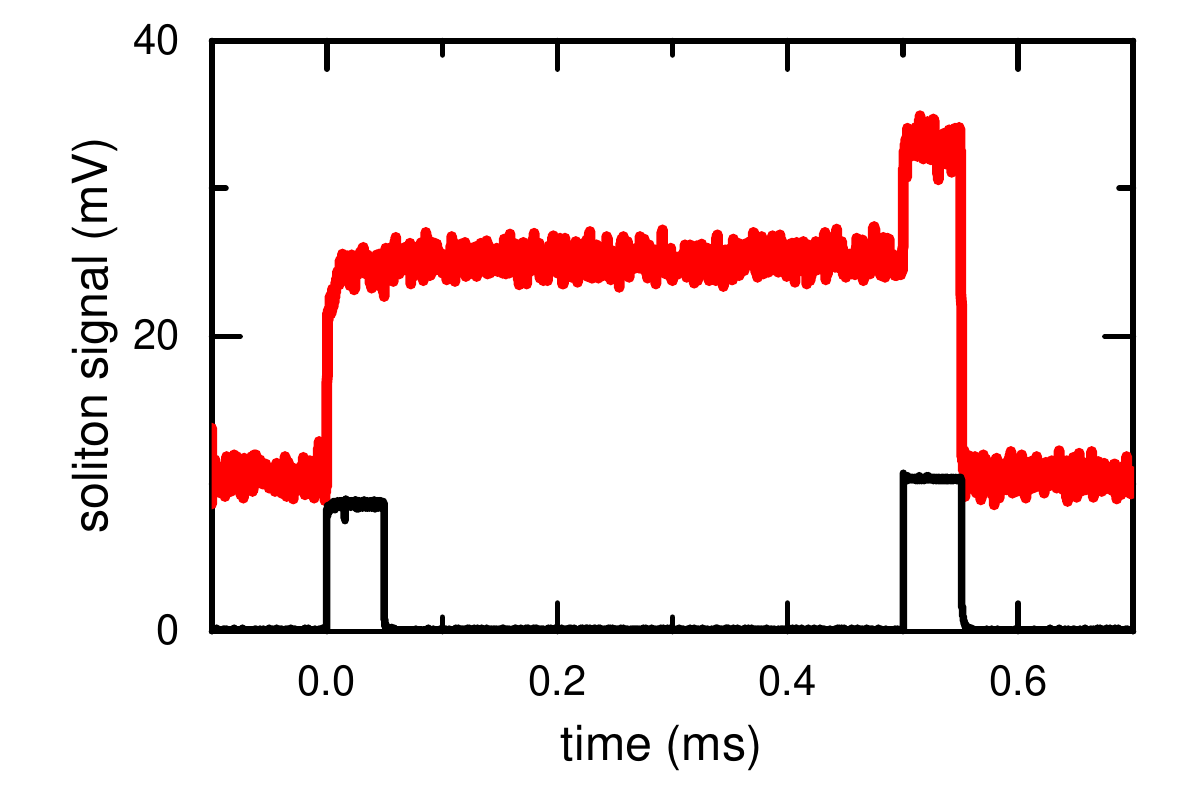}
\sidecaption\caption{\label{fig:flip-flop} Successful flip-flop event. The black trace shows the two input pulses (not to scale), the red trace the response. The red trace has been smoothed over 1~$\mu$s to remove high frequency noise. The offset of 10~mV is an electronic bias.    Temperature of VCSEL submount:  47$^\circ$C.}
\end{figure}

\begin{figure}[hbt]
\includegraphics[width=77mm]{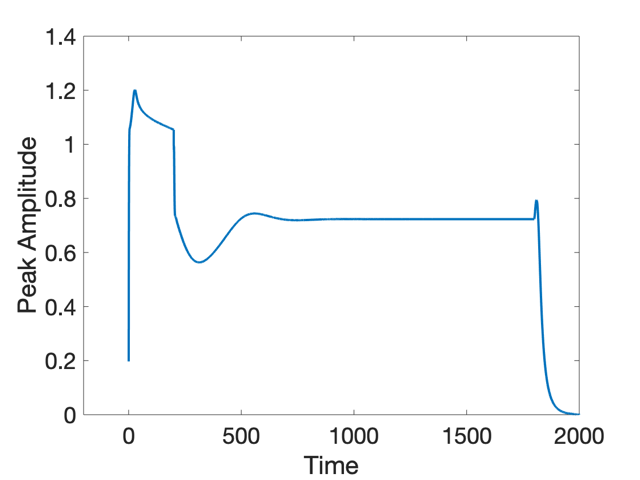}
\sidecaption\caption{\label{fig:sim} Amplitude at centre of soliton during a numerically simulated flip-flop sequence. Optical injection takes place from $t=0$ to $t=200$ (amplitude 0.3), current injection from $t=1800$ to $t=2000$ (amplitude 0.7).   Parameters (see \cite{jimenez16}): $\Theta = -2.3$, $\alpha=5$, $\gamma=0.01$, $\lambda=0.0271$, $\sigma=0.6$ and $\mu=0.55$. Time unit is the inverse of the cavity lifetime, about 10~ps. }
\end{figure}

A simulation demonstrating flip-flop operation of fundamental VCSEL solitons was performed using the model presented in \cite{jimenez16} (a simplified version, without the external-cavity delay, of the model \cite{scroggie09} used to reproduce the switch-on events in \cite{radwell10}) and is displayed in Fig.~\ref{fig:sim}. No detailed investigation or matching of parameters is intended at this point, just a qualitative demonstration. Similar to Fig.~\ref{fig:flip-flop}, the sequence is initiated by an optical pulse close to the soliton frequency at $t=0$. This causes a strong burst in the intra-cavity field, followed by a sudden drop as the end of the IB. Then the system approaches the steady-state in an oscillatory fashion. This kind of transient is typical of relaxation oscillations in semiconductor lasers but is not resolved in the current experiment as it takes only a few nanoseconds in the simulations. It should be noted that the actual switch-on transients observed experimentally and reproduced numerically in Ref.~\cite{radwell10} are even more complex than this due to the external-cavity dynamics. However, these complex transients still die out within tens of nanoseconds. At the end (at $t=1800$), a pulse is added to the current parameter to simulate the incoherent optical pumping. This is followed by a rapid switch-down of the soliton.

Obviously, more work is needed to explore and to optimize parameters but the results presented here are a proof-of-principle of flip-flop operation of laser cavity solitons via the carriers and not via parasitic thermal or non-local effects in both experiment and theory.

\section{Conclusions and Outlook}\label{sec:conclusion}
We demonstrated in this contribution vector vortex solitons with spatially inhomogeneous polarization forming spontaneously from symmetry breaking. The current experiment uses the full vector properties of light except helicity yielding a beautiful connection between nonlinear science and singular optics (or structured light). Another motivation to investigate optical solitons, in particular dissipative solitons, is their use in all-optical processing and memory applications. We reported on the flip-flop operation of laser cavity solitons via an in principle all-electronic process. Although it seems that applications of spatial solitons for parallel information storage are limited due to the sensitivity to detuning fluctuations, using a VCSEL (or, for power scaling, an optically pumped semiconductor disk laser) to create VVB might be a simple and cost-effective option. In particular, such a system might be able to switch between different VVB on time scales of hundreds of nanoseconds in response to internal parameter changes or external stimulus, whereas spatial light modulators have typically responses on the millisecond time scale. We note that not only the time scale for flip-flop operation of fundamental solitons can be improved by parameter optimization, but that switching between different vortex states might be faster as it does not involve a significant change of $S_0$ and hence carrier number in the semiconductor. This adaptability might be useful in polarization modulated spectroscopy, quantum optics or sensing applications. Obviously, some work is needed to achieve robustness in addition to flexibility.

Apart from potential applications, it is fascinating to think about the options of using external stimuli for switching between VVB for fundamental investigations. We floated the idea before that maybe VBB states we do not observe in an experimental situation might be still stable, but not accessible via the conventional LI-curves. This could be investigated by the injection of `structured light'. By coupling the cavity to a spatial light modulator \cite{naidoo16} it would be also possible to change the anisotropies in a controlled way and thus to understand the selection between different VVB. Another option is to use controlled strain to change the intrinsic anisotropies of the VCSEL \cite{doorn98,panajotov00,lindemann19}, although the bending methods used in \cite{panajotov00,lindemann19} are probably not directly suitable for broad-area VCSELs. As the information on the polarization direction is not expected to be stored directly in the semiconductor electron-hole plasma, memory and hysteresis effects are likely to be related to differences in detuning conditions for nearly degenerate VBB. It will be interesting to include the spin degrees of freedom in the consideration as some memory of them is stored in the semiconductor and found not only to be important to understand some aspects of polarization switching in VCSELs \cite{sanmiguel95b,exter98b,sanmiguel00a,ackemann05b} but also enabling novel spintronic applications \cite{lindemann19}. There are indications that the helicity becomes nonzero and spatially varying in the VCSEL with frequency-selective feedback \cite{jimenez17} but systematic investigations are lacking.

In summary, the results arguably support the notion of nonlinear optics being an excellent workhorse for investigations on conservative as well as dissipative solitons. Optical systems are not only highly controllable in parameters and have reasonably  fast time scales to allow repeated experiments, but also - via the polarization degrees of freedom -- many degrees of freedom to play with, provide an easy interface for outside control via beams and the options to tailor settings via feedback, coupled cavities, spatial light modulators and polarization changing elements.

\begin{acknowledgement}
We are grateful to Jesus Jimenez-Garcia and Pedro Rodriguez for the collaboration in the earlier stages of the investigation of the vector vortex beams \cite{jimenez17}.  We note in particular that it was Jesus Jimenez-Garcia who realized that some of the states he observed had very peculiar polarization properties and worked out a first connection to VBB. The sabbatical of T. Guillet at Strathclyde was supported by the CNRS.
\end{acknowledgement}
%
%\bibliographystyle{spphys}
%\bibliography{ThorstenLit}
%\input{references}

\end{document}